# Neural Generation of Blocks for Video Coding

**Jonah Probell**
SoundHound
`jonah@probell.com`

## Abstract

Well-trained generative neural networks (GNN) are very efficient at compressing visual information for static images in their learned parameters but not as efficient as inter- and intra-prediction for most video content. However, for content entering a frame, such as during panning or zooming out, and content with curves, irregular shapes, or fine detail, generation by a GNN can give better compression efficiency (lower rate-distortion). This paper proposes encoding content-specific learned parameters of a GNN within a video bitstream at specific times and using the GNN to generate content for specific ranges of blocks and frames. The blocks to generate are just the ones for which generation gives more efficient compression than inter- or intra- prediction. This approach maximizes the usefulness of the information contained in the learned parameters.

## 1. Introduction

Many attempts have been made to improve upon the H.264 AVC [1], H.265 HEVC [2], and AV-1 [3] video coding standards using neural network technology. Few promise improved compression efficiency across the wide range of content and practical implementations of decoders for all applications that such standards serve.

Within a given architecture (number of nodes, number of layers, activation functions, recurrences, etc.) of a trained generative neural network (GNN), its learned parameters (weight and bias values) contain compressed information. Training is a form of data compression. For example, a generative neural network trained using generative adversarial networks (GAN) [4] for generating images of fake celebrity faces [5] contains information about how faces look. An appropriately trained network can produce a high-resolution image of a face from a given set of network parameters and an input vector that, together, use far less data than the output image that it can produce. This principle can be applied to future digital video compression codecs to improve decoders' compression efficiency (rate-distortion) and energy consumption.

Prior work applying neural networks to still image compression shows significant improvements over conventional compression standards [6]. Prior work applying neural network techniques to video compression have focused on processing full video frames [7] [8] [9], object recognition [10], modeling motion and residuals in a latent space [11], improvements to bi-prediction [12], postprocessing images [13], or integrated convolutional neural networks to learn residuals and various other techniques [14]. Prior work has also focused on fixed neural networks to apply to any bitstream. This paper proposes training a GNN on a specific subset of blocks as part of encoding, including the GNN parameters in the bitstream, and applying the GNN to generate content as part of decoding.

This paper's technique has two main differences from prior work. First, generation reoccurs for subsets of blocks in a video sequence. Second, inference need not produce reasonable results for inputs unseen during training. In other words, the network need not generalize. Therefore, training needs no regularization, and overfitting is desirable.

Below, references to blocks of video encompass blocks and macroblocks, including non-square blocks, equivalent blocks in each or every color component, and blocks of interleaved content.

## 2. Method

### 2.1 Generative networks

For a given image, it is possible to train a set of parameters for a GNN that, when evaluated through multiplications, additions, and activation functions, generate an output that matches the image with low loss. No input to the GNN is needed, as it contains, within its parameters, the information representing the image. Conceptually, the input to any node at the lowest layer of the neural network is a defined constant such as 1.

Training a GNN can go through many iterations to improve the parameters to reproduce the image with lower loss. The larger the GNN architecture, the lower the achievable loss. Note that there is only one correct output. It is not necessary to train for inference or generation on yet-unseen inputs. So, training adversarially against a discriminator does not apply. Furthermore, training techniques needed to avoid overfitting are unneeded.

### 2.2 Block generation as an alternative to prediction

The inter- and intra- prediction techniques of state-of-the-art codecs provide good compression efficiency for most blocks of most videos. However, inter-prediction efficiency is low for content with some kinds of motion such as panning, in which case new content appears at one or two adjacent frame edges, zooming out, in which case new content appears at all four frame edges, or zooming in, in which case visual detail increases throughout the frame. Where content appears, there is no reference for prediction. Where detail increases, the best reference has high error. So, the residuals contain most of the information for such content. Residuals only represent magnitudes of non-varying functions (sine waves) at wavelengths equal or lower than a block size. They have no way to represent irregular features or features larger than a block. As a result, block-wise prediction cannot capture certain redundancies in content. Therefore, residuals have more energy than theoretically necessary and lower than optimal compression through quantization.

Intra-prediction efficiency is low for frames with fine detail and, especially, curves and irregular shapes.

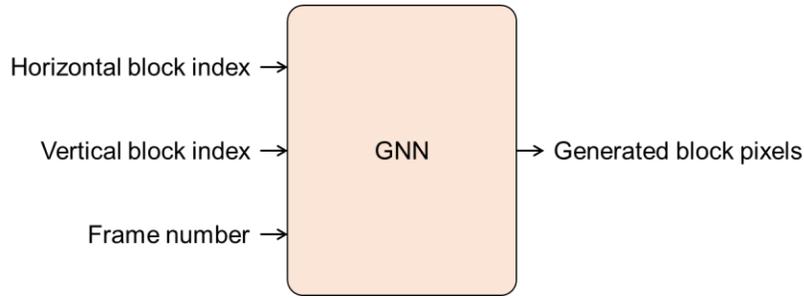

Figure 1: A GNN for video block generation

Neural networks can represent features of irregularly shaped content at different levels of detail with high compression. Therefore, it is more efficient to encode blocks for which the best prediction reference would have low accuracy using parameters for a GNN that can generate the block content. Where inter- or intra-prediction can give high prediction accuracy, block generation has no benefit. By generating only blocks for which doing so gives a significant compression improvement over prediction, the information that the GNN learns to store is allocated to just the content just where it can have the most benefit.

Content generation, though trained for minimal loss, is imperfect. Therefore, differences between source and generated content should be encoded as residuals much as with predicted blocks.

## 2.3 GNN function

Although a GNN for predicting blocks, theoretically, needs no input to generate a large number of outputs, in practice, it is reasonable to generate a block-sized output for any given input of block horizontal and vertical indices and a frame number.

Figure 1 shows a GNN with its inputs and output. A 4K UHD resolution (3840 x 2160 pixels) frame has 120 horizontal and 68 vertical block indices. A codec may set a maximum number of frames over which to use a GNN parameter set before a new parameter set is encoded. 120 frames (2 seconds at 60 fps) might be a reasonable maximum for most consumer video applications. Coordinating the encoding of parameter sets with keyframes is generally desirable since it avoids having to learn two content sets across scene changes, which would be inefficient, and because the GNN parameters are necessary for syncing decoding during random entry into a stream sequence.

A codec standard may define the output of a GNN to be a fixed size pixel block such as 32x32 (1024-pixel output). Coincidentally, this is the size of images in the CIFAR-10 dataset [15], and the down-sampled version of the ImageNet data set [16], both of which many researchers have used to train for other neural image generation tasks. The task of a GNN for a video codec is comparable to generating images matching an embedded label space learned for specific video content, the label being the block horizontal and vertical coordinates and frame number.

Smaller size block generation may require more overhead for selection between generation and prediction, as discussed below. Larger size output block generation may lose efficiency by generating some content that could have been more accurately inter- or intra-predicted just because it is adjacent to content that benefits from generation.

It is also possible to define a codec to support the generation of rectangular blocks that support quad-tree-based block partitioning. Furthermore, it is possible to train a GNN to predict small blocks that only use a portion of a larger number of GNN outputs. It would be reasonable to define the outputs used by their sub-block horizontal and vertical offsets within a larger block output size.

## 2.4 GNN parameters in the bitstream

A GNN can learn parameters that are especially efficient for a set of spatially and/or temporally neighboring blocks. Covering neighboring blocks takes advantage of a neural network's ability to represent features at both high and low levels of detail. For a conceptual example, a face moving into frame will have features with spacing and orientation that are face-like across neighboring blocks and the same blocks within neighboring frames. Naturally, training does not identify face-like features. It learns whatever features are recognizable within whichever blocks of whatever content the GNN generates.

The trained GNN parameters are encoded in the bitstream. Training may be done across multiple frames to benefit from temporal redundancy. At specific times, such as keyframes, periodically, or at the encoder's discretion, a new set of GNN parameters is encoded in the bitstream. The frequency of encoding new GNN parameter sets is a trade-off between bits used for GNN parameters and generation accuracy. Reusing a GNN parameter set over a very long number of frames will not benefit from the continued presence of features across chronologically sequential frames. For mass consumer video, re-encoding a GNN parameter set every 8 to 60 frames might provide optimal encoding efficiency. Larger GNN architectures and more thorough training allow for less frequent transmission of GNN parameter sets.

## 2.5 Controlling overhead

To encode whether to predict or generate a block requires 1 bit of data. Doing so individually for each block would create an overhead cost to encoding that is unacceptable for low bitrate applications. A codec (or profile of a codec) could support one or more approaches to minimizing overhead.

One approach is to allow, on a per-frame basis, defining one or several rectangular regions in which blocks are generated or regions in which they are not generated. Defining a rectangle only requires two sets of a horizontal and vertical block number. Encoding this requires little overhead but could inefficiently force generation of blocks that could have been predicted with high accuracy and, thereby, waste the information content of GNN parameter sets. On the other hand, regions of low prediction efficiency tend to be regions of one or several block margins at edges of frames when video has panning or zooming out. Those margins regions are rectangles that will tend to have a high portion of blocks for which generation is more accurate than prediction.

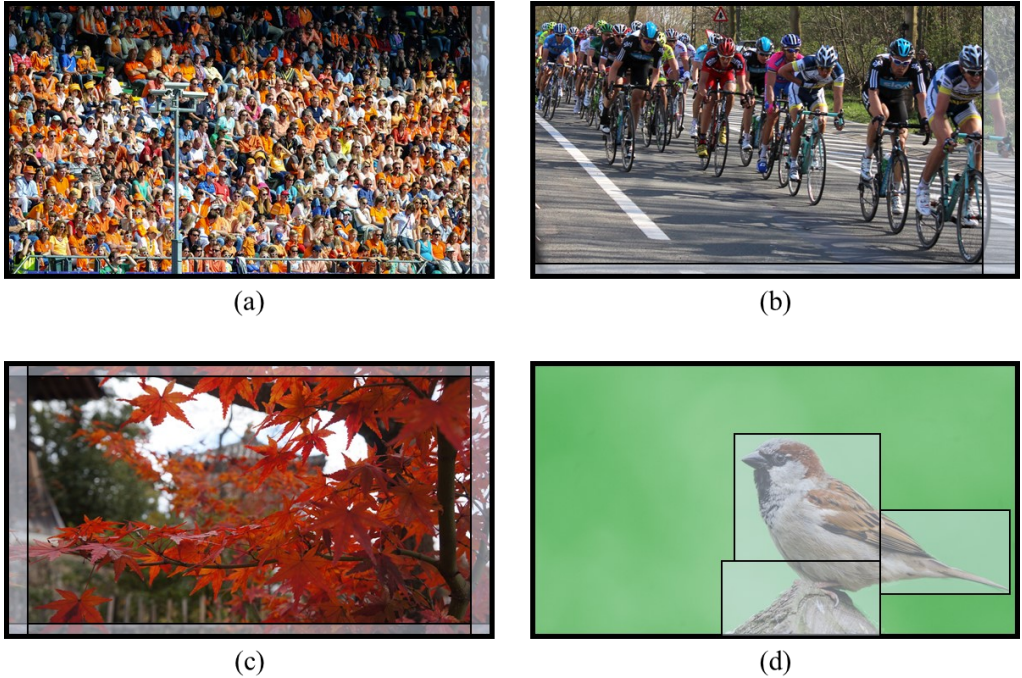

Figure 2: Regions for block generation in a frame with (a) rightward panning, (b) right and downward panning at different rates, (c) zooming out, and (d) zooming in with high-detail regions.

Figure 2 shows an example (a) of rightward panning in which content of people is generated within a region on the right edge of the frame. The optimal region width varies with the speed of panning. Another example (b) shows two-axis panning rightward and downward. The right edge region width is larger than the bottom edge region height because the horizontal panning is faster than the vertical panning. Another example (c) shows zooming out from the center of the frame, in which case the left and right edge region widths compared to the top and bottom edge region heights are proportional to the frame aspect ratio. Another example (d) shows a scenario of zooming in in which the encoder has selected regions of fine detail for generation, leaving regions of low detail for prediction.

Another approach is to allow, on a per-frame basis, defining one or several rectangular regions of blocks in which blocks have a selection bit. Not necessarily all blocks in the region are generated. Instead, the encoding has a bit to select between generation and prediction for each block in the region. Per-block selection gives the fine control of per-block generation but limits the overhead cost of selection bits to just certain ranges of blocks. Global motion estimation can inform an encoder's selection of what edges and how many blocks on each edge can benefit from region-based constraints on selection.

Another technique for controlling block overhead is to choose the frames in which to send new GNN parameter sets. Using a parameter set across scene changes would be inefficient. A simple approach would be to send GNN parameter sets with each keyframe or each I frame, the parameters being for use across all frames until the next frame carrying a new set.

Another approach is to allow encoders to choose when to send a parameter set. Having a choice allows the encoder to optimize coding efficiency by choosing the generation rate based on content. Content with high motion or fine detail will benefit from more frequent updates of the GNN. Leaving flexibility to the encoder allows for better optimization based on bit rate requirements with little additional cost to decoder design.

A related technique is to gradually encode GNN parameters for the next update throughout a sequence of frames to avoid bandwidth bursts at the frames that first use each GNN parameter sets.

## 2.6 Architecture definition

The GNN architecture affects decoder requirements. A video codec standard might define a GNN architecture (number of nodes, number of layers, activation functions, recurrences, etc.). It would be possible to have a different architecture per profile to fit the requirements of different applications. Real-time profile encoders, for example, typically would not have enough time to learn optimal GNN parameters, and so they might omit generation altogether.

It is also possible to define several GNN architectures and allow an encoder to select between them based on content and target bitrates. Decoders would need to support up to the most complex architecture type allowed for its codec profile.

Another approach is to allow the encoder to specify the architecture within the bitstream. Microsoft Net# [17] or the Keras API [18] are examples of existing languages to specify neural network architectures. Providing the encoder flexibility to define the GNN architecture can further improve compression efficiency by requiring a minimal number of bits to carry the information needed to generate the most critical blocks.

### 2.6.1 GNN Architectures

The field of neural image generation is improving rapidly. The best architectural approach for a video coding standard will require more experimentation and consensus of industry participants.

Consider a use case of 4k (3840x2160 pixels or 120x68 blocks of 32x32 pixels) at 30 fps encoded at 40 Mbps having 1 parameter set per second. If the encoder chose, for a high motion scene, to predict an average of 50% of the blocks on each of a horizontal edge and a vertical edge (94 blocks) for each frame, it would require 2820 GNN calls by the decoder to generate 1.2% of decoded blocks.

A GNN for 32x32 pixel blocks (1024 pixels) with 4:2:0 chroma subsampling has 1536 outputs generated from 3 inputs (block x, block y, and frame number). A GNN architecture with approximate 100,000 parameters at 10 bits per parameter consumes a reasonable 2.5% of the bandwidth (1M bps from 40 Mbps). The amount of marginal compression efficiency that results from allocating 2.5% of bandwidth to improve the bitrate of the worst 1.2% of blocks depends on the histogram of bitrate allocation to blocks. A high variance distribution, as occurs for high compression ratios, will see more benefits from generation.

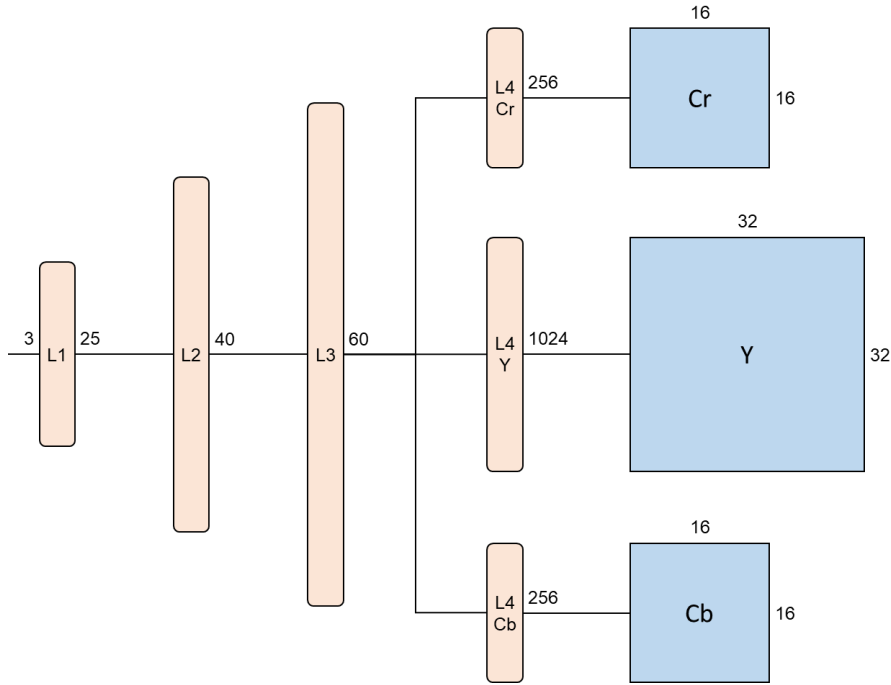

Figure 3: A proposed 93,696 parameter GNN architecture

Figure 3 shows an example of such a GNN architecture. It has a first layer (L1) with 25 activations from the 3 inputs (100 parameters including weights and biases). It tends to learn the locations of instances of complex irregular objects that can span many blocks such as close human faces, animals, vehicles, buildings, text, etc. A second layer (L2) has 40 activations from L1 (1040 parameters). It tends to learn block-scale irregular features and relative locations common to instances of complex objects. A third layer (L3) has 60 activations from L2 (2460 parameters). It tends to learn sub-block sized fine features such as textures and offsets of complex features within the rectangular blocks. All such features are useful for prediction across all color components. A fourth layer (L4) produces the pixel output values for each of the 1536 Y, Cr, and Cb color channels (93,696 parameters). The total GNN architecture has 97,296 parameters.

Because of the limited number of calls of the GNN and commonality of features between video frames, a simple fully-connected deep neural network (DNN) architecture, as shown in Figure 3, might be appropriate. However, a sequence of 2-dimensional bi-direction LSTMs, as used in the PixelRNN of van den Oord et al. [19], can yield improvements by taking advantage of the long-distance spatial dependencies between objects and feature types. Furthermore, a sequence of Resnet [20] or standard CNN blocks with upsampling, as used in the image GNNs of BigGAN by Lucic et al. [21] and large scale GANs by Brock et al. [22] might be appropriate, especially since learned convolutional kernels tend to be reusable temporally across frames. A standard-defined architecture could adopt a combination of elements of CNN, LSTM, and DNN nodes.

ReLU is a suitable activation function for neural image generation applications. Its always-positive result is right for pixel values. Furthermore, it does not suffer from the vanishing gradient problem of deep networks with sigmoid activations. The slow training drawback of ReLU is of little concern for video block generation because there is only a finite, small (typically hundreds to tens of thousands) number of data samples to learn. Most importantly, ReLU requires no trigonometric functions or lookup tables, making it straightforward to implement in hardwired and cost- and power-sensitive video decoder applications.

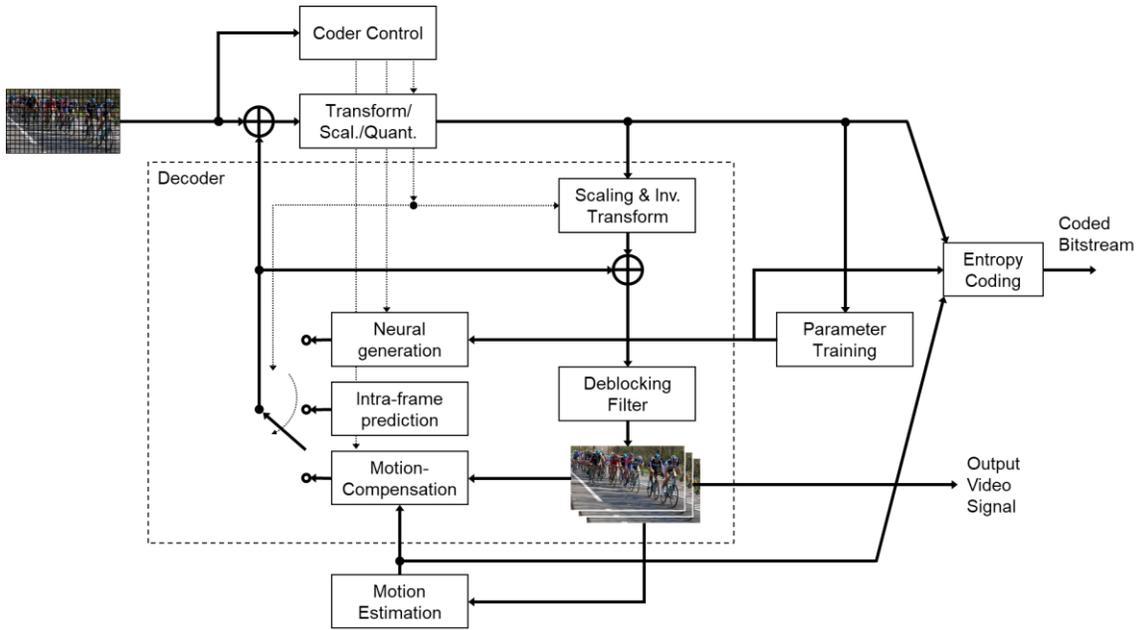

Figure 4: Typical video encoder with neural generation

## 2.7  Encoder techniques

Figure 4 shows a block diagram of a typical video encoder with neural generation. It has the coder control, transforms, scaling, quantization, in-loop deblocking filter, intra-prediction, inter-frame motion compensation, motion estimation, and entropy coding as in typical conventional video encoders. [2] However, it also has a coder control-selectable neural generation function that the encoder can select on a per-block basis instead of intra- or inter-prediction. Outside of the decoder group is a parameter training function that is part of the encoding process. It learns the parameters needed for the neural generation function in the decoder.

A GNN training loss function ideally considers a fully compressed bit count and/or PSNR across all frames that use a given parameter set. However, performing conventional encoding optimizations such as block partitioning, intra-prediction modes or inter-prediction motion vectors, entropy coding, and filter coefficient selection for each training iteration would be prohibitively expensive for all but the most cost-insensitive applications. Instead, selecting blocks to generate and the resulting GNN training can be a sub-routines of a conventional rate-distortion optimization process. To trade-off training effort against optimization, an encoder can configure the algorithm for choosing ranges in which to generate blocks, the number of block generations to try within the ranges, the different GNN architecture configurations, and, for each of the above, the number of parameter training iterations.

### 2.7.1 Parameter training

For any given architecture configuration, the training process may be subject to various hyperparameters such as learning rate, momentum, and mini-batch size. A choice of optimization algorithm, such as stochastic gradient descent (SGD) or Adam, is also needed and worth further experimentation.

Since overfitting to a defined set of possible outputs is desirable, validation and regularization techniques such as dropout do not apply. Likewise, the concept of separate training, test, and validation sets does not apply.

### 2.7.2 Architecture configuration

A codec standard may define a specific GNN architecture or set of GNN architectures. However, if a standard leaves the choice of architecture to the encoder, training can remove nodes that it finds to provide little contribution to output accuracy and, thereby, further improve coding efficiency.

Of course, allowing an encoder to choose any arbitrary GNN architecture would make the decoding task unbounded. A standard that allows an encoder configurable GNN architecture should also set maximums on nodes and/or layers to set practical limits on decoder size, cost, and power consumption.

### 2.7.3 Parallelization

Two basic approaches are possible for an encoder that performs neural generation of blocks. One approach applies neural network training as a step within a rate-distortion-minimizing predictive encoder. The other approach performs prediction encoding as a step within computing the loss function of GNN training. Many variations of speculative optimization decision making are possible. Again, further research is needed into encoder algorithms.

Regardless of which approach, the innermost loop may run in parallel with different configurations in a multi-processing environment. A cloud-server based encoder may use massive parallelism to optimize encoding. Compared to conventional codecs, a codec with neural block generation creates much more complexity to compute the optimal encoding. However, this has the benefit of creating more opportunities for parallelism, enabling the efficient application of modern massive parallel computing environments to video bitstream optimization than is possible for encoding for conventional codecs.

## 3. Discussion and further research

It is interesting to note that, if a training loss function depends on the fully compressed bit count (or at least the bits for encoding residuals), training will learn to adapt to the frequency component weights of any given quantization matrix. In other words, the GNN will focus its learning on the information most important for psycho-visually accurate reconstruction.

Application-specific encoder optimizations need further research. Many approaches to architecture and training recipes are possible. The ground is fertile for future papers.

Optimized chip designs and cores for video decode using neural generation of blocks can help to develop an ecosystem that supports consumer adoption of such a standard. Fortunately, well-known techniques for optimizing the multiply-add performance of processors for common digital signal processing algorithms are generally applicable to efficient performance on the high-throughput multiply-adds needed for neural network nodes.

Video coding using neural block generation is eminently useful for encoding relatively small libraries of premium content such as the video libraries of Netflix, Amazon, and Disney. Such services compete to deliver the highest possible quality of high-resolution video to consumers over potentially low bandwidth home internet connection. Such services have the financial resources for extreme amounts of processing power to achieve high-efficiency encodings.

Financially well-resourced video providers with vast libraries of long-tail content, such as YouTube, can also benefit from video coding with neural block generation. They can allocate server resources to videos in proportion to their popularity. Doing so provides the best possible aggregate bandwidth savings. It also provides the best compression efficiency to the most popular content, giving the best possible aggregate user experience.

Since training a neural network is very compute-intensive and time-consuming, the GNN option within a standard is likely to be unused in real-time applications such as video conferencing or safety-critical cameras. Furthermore, the large amount of computation required for GNN training is unlikely to be useful in power-sensitive encoding applications such as mobile phone cameras. However, its compression-efficiency improvements can save power for decoders in mobile devices by minimizing radio usage.

## 4. Conclusion

A neural network specifically trained to generate content for a limited number of blocks within a video sequence provides efficient compression of image features within the predicted blocks. Applying the technique specifically to blocks for which intra- or inter-prediction would have large residual error improves overall compression efficiency (rate-distortion). Training such a neural network is much simpler than training ones for generalized inference. Neural network generation can use existing DSP compute resources efficiently. Therefore, neural generation of blocks for video coding offers an improvement over state-of-the-art video codecs that is worth inclusion in future video coding standards.

## 5. References

[1] T. Wiegand, G. J. Sullivan, G. Bjontegaard and A. Luthra, "Overview of the H. 264/AVC video coding standard," *IEEE Transactions on circuits and systems for video technology,* vol. 13, no. 7, pp. 560-576, 2003.

[2] G. J. Sullivan, J.-R. Ohm, W.-J. Han and T. Wiegand, "Overview of the high efficiency video coding (HEVC) standard," *IEEE Transactions on circuits and systems for video technology,* vol. 22, no. 12, pp. 1649-1668, 2012.

[3] Y. Chen, D. Murherjee, J. Han, A. Grange, Y. Xu, Z. Liu, S. Parker, C. Chen, H. Su, U. Joshi, C.-H. Chiang, Y. Wang, P. Wilkins, J. Bankoski, L. Trudeau, N. Egge, J.-M. Valin, T. Davies, S. Midtskogen, A. Norkin and P. de Rivaz, "An overview of core coding tools in the AV1 video codec," *2018 Picture Coding Symposium (PCS),* pp. 41-45, 2018.

[4] I. Goodfellow, J. Pouget-Abadie, M. Mirza, B. Xu, D. Warde-Farley, S. Ozair, A. Courville and Y. Bengio, "Generative adversarial nets," *Advances in neural information processing systems,* pp. 2672-2680, 2014.

[5] S. Sk, J. Jabez and V. M. Anu, "The Power of Deep Learning Models: Applications," *International Journal of Recent Technology and Engineering (IJRTE),* vol. 8, no. 2S11, 2019.

[6] G. Toderici, D. Vincent, N. Johnston, S. J. Hwang, D. Minnen, J. Shor and M. Covell, "Full Resolution Image Compression with Recurrent Neural Networks," in *Proceedings of the IEEE Conference on Computer Vision and Pattern Recognition*.

[7] E. Gelenbe, M. Sungur, C. Cramer and P. Gelenbe, "Traffic and video quality with adaptive neural compression," *Multimedia Systems,* pp. 357-369, 1996.

[8] Z. Chen, T. He, X. Jin and F. Wu, "Learning for Video Compression," *IEEE TRANSACTIONS ON CIRCUITS AND SYSTEMS FOR VIDEO TECHNOLOGY,* 09 01 2019.

[9] J. Han, S. Lombardo, C. Schroers and S. Mandt, "Deep Generative Video Compression," in *33rd Conference on Neural Information Processing Systems*, Vancouver, Canada, 2019.

[10] S. Ghorbel, M. Ben Jemaa and M. Chtourou, "Object-based Video compression using neural networks," *International Journal of Computer Science,* vol. 8, no. 4, pp. 139-148, 2011.

[11] A. Djelouah, J. Campos, S. Schaub-Meyer and C. Schroers, "Neural Inter-Frame Compression for Video Coding," in *IEEE International Conference on Computer Vision*, 2019.

[12] Z. Zhao, S. Wang, S. Wang, X. Zheng, S. Ma and J. Yang, "Enhanced Bi-Prediction With Convolutional Neural Network for High-Efficiency Video Coding," *IEEE TRANSACTIONS ON CIRCUITS AND SYSTEMS FOR VIDEO TECHNOLOGY,,* vol. 29, no. 11, pp. 3291-3301, 11 11 2019.

[13] G. Lu, W. Ouyang, D. Xu, Z. Zhang, Z. Gao and M.-T. Sun, "Deep Kalman Filtering Network for Video Compression Artifact Reduction," *European Conference on Computer Vision,* 2018.

[14] S. Ma, X. Zhang, C. Jia, Z. Zhao, S. Wang and S. Wang, "Image and Video Compression with Neural Networks: A Review," *IEEE TRANSACTIONS ON CIRCUITS AND SYSTEMS FOR VIDEO TECHNOLOGY,* 10 04 2019.

[15] A. Krizhevsky, "The CIFAR-10 dataset," Canadian Institute For Advanced Research, [Online]. Available: https://www.cs.toronto.edu/~kriz/cifar.html. [Accessed 23 08 2020].

[16] "Downsampled ImageNet," Stanford Vision Lab, Stanford University and Princeton University, [Online]. Available: http://www.image-net.org/small/download.php. [Accessed 23 08 2020].

[17] Microsoft, "Guide to Net# neural network specification language for Machine Learning Studio (classic)," 01 03 2018. [Online]. Available: https://docs.microsoft.com/en-us/azure/machine-learning/studio/azure-ml-netsharp-reference-guide. [Accessed 15 08 2020].

[18] Keras.io, "Keras API reference," [Online]. Available: https://keras.io/api/. [Accessed 15 08 2020].

[19] A. van den Oord, N. Kalchbrenner and K. Kavukcuoglu, "Pixel Recurrent Neural Networks," vol. 1601.06759v3, 19 08 2016.

[20] K. He, X. Zhang, S. Ren and J. Sun, "Deep Residual Learning for Image Recognition," in *Proceedings of the IEEE conference on computer vision and pattern recognition*, 2019.

[21] M. Lucic, M. Tschannen, M. Ritter, X. Zhai, O. Bachem and S. Gelly, "High-Fidelity Image GenerationWith Fewer Labels," 14 05 2019.

[22] A. Brock, J. Donahue and K. Simonyan, "Large scale gan training for high fidelity natural image synthesis," *ICLR 2019,* no. 1809.11096v2, 25 02 2018.